# Electron-Phonon Scattering in Semiconductor Structures with One-Dimensional Electron Gas

**D.V. Pozdnyakov**[1], **V.O. Galenchik**

*Radiophysics and Electronics Department, Belarus State University,*
*Nezavisimosty av. 4, 220050 Minsk, Belarus*

**Abstract**
In this study a method for calculation of the electron-phonon scattering rate in semiconductor structures with one-dimensional electron gas is developed. The energy uncertainty of electrons is taken into account.

At present both theoretical and experimental investigations of semiconductor structures with one-dimensional (1D) electron gas are of great interest [1–3]. Because of the application of these structures in electronics, understanding of their electrophysical properties becomes of prime importance. The Monte Carlo simulation is one of the most promising approaches in the study of electron transport in the systems with 1D electron gas [4, 5]. It is known that the correct application of this simulation technique is only possible if the correct and comprehensive description of charge carrier scattering processes is available. In Refs. [6, 7] we have described the electron energy uncertainty effect on the acoustic and polar optical phonon scattering rates in GaAs quantum wires using the results obtained in Ref. [8]. However, the method developed in the indicated papers was only a simplified approach to accounting for the influence of electron energy uncertainty on the scattering rate. The essence of that approach is to formally replace the equation for the ideal density of states with that taking into account the energy level broadening. To derive a more accurate description of the electron-phonon scattering rate in semiconductor structures with 1D electron gas than that developed in Refs. [6, 7] let us consider the influence of the energy uncertainty on the scattering processes in a more rigorous way.

According to Refs. [6, 7, 9] in idealized semiconductor structures with 1D electron gas the electron-phonon scattering rate from the state with quantum numbers $i = \{n, l\} = 1, 2, ..$ and energy $E_i = E_{nl}$ to the state with the numbers $j = \{n', l'\} = 1, 2, ..$ and energy $E_j = E_{n'l'}$ can be presented as

$$W_{ij}^{\alpha\beta}(E) = \frac{L}{\hbar} \int_0^\infty \left|M_{ij}^{\alpha\beta}(E,\varepsilon)\right|^2 \left(1 - f_j^\beta(\varepsilon)\right) k'(\varepsilon) \delta\left(\varepsilon + \Delta E_{ij}^{\alpha\beta} - E\right) d\varepsilon =$$

$$\frac{L}{\hbar} \left|M_{ij}^{\alpha\beta}\left(E, E - \Delta E_{ij}^{\alpha\beta}\right)\right|^2 \left(1 - f_j^\beta\left(E - \Delta E_{ij}^{\alpha\beta}\right)\right) k'\left(E - \Delta E_{ij}^{\alpha\beta}\right) \Theta\left(E - \Delta E_{ij}^{\alpha\beta}\right), \tag{1}$$

where $\hbar$ is the Planck constant, $L$ is the quantum wire length, $E$ is the electron kinetic energy, $\alpha$ is the superscript distinguishing emission or absorption, $\beta$ is the superscript which labels the forward or backward scattering, $\Delta E_{ij}^{\alpha\beta} = E_j - E_i \pm E_{ph}$ is the difference between the final and initial state energies when a phonon with energy $E_{ph}$ is emitted or absorbed, $\delta$ is the Dirac $\delta$-function, $f_j$ is the kinetic energy distribution function of electrons in the final state, $M_{ij}$ is the scattering matrix element, $\Theta$ is the unit step function, $k$ is the electron wave vector absolute value, and $k'(\varepsilon) = \partial k/\partial\varepsilon$ ($k'(\varepsilon_0) = (\partial k/\partial\varepsilon)|_{\varepsilon=\varepsilon_0}$).

This formula describes individual scattering acts in the two-particle approximation [9–12]. According to Refs. [9–12], the two-particle approximation is adequate in most of practically es-sential cases and correctly describes the scattering processes in weakly perturbed quantum systems. At the same time, the neglect of multiple scattering acts yields wrong results in the scattering theory for systems with 1D electron gas [13, 14]. Therefore to take into account the multiple scattering acts it is necessary to use the following more general equation [9–11] instead of Eq. (1):

$$W_{ij}^{\alpha\beta}(\langle E \rangle, \Gamma_i) = \frac{L}{\hbar} \Theta(\langle E \rangle - \Delta E_{ij}^{\alpha\beta}) \times$$

$$\int_0^\infty \left| M_{ij}^{\alpha\beta}(\langle E \rangle, \varepsilon) \right|^2 (1 - f_j^\beta(\varepsilon)) k'(\varepsilon) \rho(\langle E \rangle, \Gamma_i, \varepsilon) d\varepsilon, \quad (2)$$

where $\langle E \rangle$ is the average electron kinetic energy, $\rho$ is the spectral (power) function, and $\Gamma_i$ is the energy characterizing the width of $\rho$. Taking into account that the probability for the particles to stay in the initial state decreases exponentially with time [9–11], the following equation for the spectral function can be derived [10, 11]:

$$\rho(\langle E \rangle, \Gamma_i, \varepsilon) = \frac{2}{\pi} \cdot \frac{\Gamma_i}{\Gamma_i^2 + 4(\varepsilon + \Delta E_{ij}^{\alpha\beta} - \langle E \rangle)^2} +$$

$$\frac{2}{\pi} \cdot \frac{\Gamma_i}{\Gamma_i^2 + 4(\varepsilon - \Delta E_{ij}^{\alpha\beta} + \langle E \rangle)^2}, \quad (3)$$

where [10]

$$\Gamma_i = \hbar \sum_{j,\alpha,\beta} W_{ij}^{\alpha\beta}(\langle E \rangle, \Gamma_i). \quad (4)$$

---

[1] pozdnyakov@bsu.by

It should be noted that in Refs. [10, 11] only the first term of Eq. (3) is used as a spectral function. However, this approximation is only valid if $(\varepsilon + \Delta E_{ij}^{\alpha\beta} - \langle E \rangle) \gg \Gamma_i$.

Thus, the calculation of electron-phonon scattering rate is reduced to solving the integral equation (4). However, this problem involves significant mathematical and computational difficulties [14]. These difficulties can be overcome by taking into account the explicit expression describing behavior of the function $\left| M_{ij}^{\alpha\beta}(\langle E \rangle, \varepsilon) \right|^2 (1 - f_j^\beta(\varepsilon)) k'(\varepsilon) \varepsilon^{1/2}$ versus $\varepsilon$. Prescribed function is quite smooth compared with $\varepsilon^{-1/2}$ in a wide range of $\varepsilon$ for various types of electron-phonon scattering processes [3, 6, 7, 9, 13, 15–17]. In this case the equation (2) can be transformed to

$$W_{ij}^{\alpha\beta}(\langle E \rangle, \Gamma_i) \approx \frac{L}{\hbar} \left| M_{ij}^{\alpha\beta}(\langle E \rangle, \langle E \rangle - \Delta E_{ij}^{\alpha\beta}) \right|^2 (1 - f_j^\beta(\langle E \rangle - \Delta E_{ij}^{\alpha\beta})) \times$$

$$\Theta(\langle E \rangle - \Delta E_{ij}^{\alpha\beta}) k'(\langle E \rangle - \Delta E_{ij}^{\alpha\beta}) \sqrt{\langle E \rangle - \Delta E_{ij}^{\alpha\beta}} \int_0^\infty \rho(\langle E \rangle, \Gamma_i, \varepsilon) \varepsilon^{-1/2} d\varepsilon =$$

$$\frac{L}{\hbar} \left| M_{ij}^{\alpha\beta}(\langle E \rangle, \langle E \rangle - \Delta E_{ij}^{\alpha\beta}) \right|^2 (1 - f_j^\beta(\langle E \rangle - \Delta E_{ij}^{\alpha\beta})) k'(\langle E \rangle - \Delta E_{ij}^{\alpha\beta}) \times$$

$$\Theta(\langle E \rangle - \Delta E_{ij}^{\alpha\beta}) \sqrt{\langle E \rangle - \Delta E_{ij}^{\alpha\beta}} \sqrt{\frac{2\Gamma_i + 2\sqrt{\Gamma_i^2 + 4(\langle E \rangle - \Delta E_{ij}^{\alpha\beta})^2}}{\Gamma_i^2 + 4(\langle E \rangle - \Delta E_{ij}^{\alpha\beta})^2}}. \quad (5)$$

Finally according to Eqs. (4) and (5) the problem of calculation of the electron-phonon scattering rate can be reduced to the solution of an algebraic equation with a polynomial of 8-th degree.

Thus, the method for calculation of the electron-phonon scattering rate in semiconductor structures containing one-dimensional electron gas is developed with the energy uncertainty of electrons taken into account.